\documentstyle[preprint,aps]{revtex}  

\begin{document}

\draft

\title{Chiral symmetry breaking in d=3 NJL model in  external
gravitational and magnetic fields}

\author{D.M. Gitman}

\address{Instituto de F\'{\i}sica, Universidade de S\~ao Paulo \\ 
Caixa Postal 66318, 05389-970-S\~ao Paulo, S.P., Brasil
}
\author{S.D. Odintsov}
\address{Dept. of Math.Analys., Tomsk Pedagogical University \\ 
634041 Tomsk, Russia,
and 
Dept. ECM, Fac. de Fisica \\ Universidad de
Barcelona, 
Diagonal 647, 08028 Barcelona, Spain}
\author{Yu.I. Shil'nov}
\address{Dept. Theor. Phys., Kharkov State University \\ 
Svobody Sq. 4, Kharkov 310077, Ukraine}

\date{\today}

\maketitle

\begin{abstract}
The phase structure of $d=3$ Nambu-Jona-Lasinio model in curved
spacetime with magnetic field is investigated in the leading order of
the $1/N$-expansion and in linear curvature approximation (an external
magnetic field is treated exactly). The possibility of the chiral
symmetry breaking under the combined action of the external
gravitational and magnetic fields is shown explicitly.  At some
circumstances the chiral symmetry may be restored due to the
compensation of the magnetic  field by the gravitational field.
\end{abstract}

$d=3$ NJL model may be very useful laboratory for the study of dynamical
symmetry breaking and composite bound states. This model is known to
be renormalizable in $1/N$-expansion and it has quite rich phase
structure. In principle, such a model may be helpful as some
high-temperature phase of the corresponding $d=4$ theory considered as
kind of the effective theory for QCD.

Recently, phase structure of $d=3$ NJL model has been investigated in
an external gravitational field \cite{b2,b3} and in an external
magnetic field \cite{b4,b5,b6,b12} (and references therein). It has been
pointed out that magnetic field (for a general introduction to the
theory in an external electromagnetic field see \cite{b7}) usually
supports the chiral symmetry breaking. On the same time, for such a
model on $S_3$ or $H_3$ spaces one can show \cite{b3} that chiral
(flavor) symmetry is always broken in $H_3$ (hyperbolic space) while
it can be restored for some values of the curvature on $S_3$
background.

Having in mind the possible cosmological applications of such
considerations \cite{b8} and the possibility of the existence of large
primordial magnetic fields at the early Universe it would be of
interest to discuss the combined effect of gravitational and magnetic
fields to phase structure. That will be the purpose of the present
letter.

We will start first from $d=3$ four-fermion model in curved spacetime:
\begin{equation}\label{1}
L=\overline{\Psi}^i \gamma^{\mu}(x)\nabla
_{\mu}\Psi^i-\frac{N\sigma^2}{2\lambda}-\sigma \overline{\Psi}^i \Psi^i\;,
\end{equation}
where $i=1,\ldots,N,\;\;N$ is number of fermion species, $\gamma^{\mu}(x)$ are
curved $\gamma$-matrices and as usually the auxiliary scalar field
$\sigma$ is introduced. We work in linear curvature approximation
(keeping only linear curvature terms in the leading correction to
effective potential) and also in $1/N$-expansion. Then it is
convenient to use the local momentum representation of the fermion
propagator in curved spacetime in the form discussed in \cite{b8}.

Integrating in the functional integral over fermion fields in the
standard way, one can get the effective potential as
following \cite{b8}:
\begin{equation}\label{2}
V(\sigma)=\frac{\sigma^2}{2\lambda}+i{\rm Sp}\ln
\left[i\gamma^{\mu}(x)\nabla_{\mu}-
\sigma\right]\;.
\end{equation}
It is convenient to work in terms of the derivative from
$V(\sigma)$. Then, using the local momentum representation of the
propagator, we get
\begin{eqnarray}\label{3}
V'(\sigma)&=&\frac{\sigma}{\lambda}-i{\rm Sp}\int \frac{d^3
k}{(2\pi)^3}\left[\frac{\hat{k}+\sigma}{k^2-\sigma^2}-
\frac{R}{12}\frac{\hat{k}+
\sigma}{(k^2-\sigma^2)^2}\right. \nonumber \\
&+&\left.\frac{2}{3}R_{\mu\nu}k^{\mu}k^{\nu}\frac{(\hat{k}+
\sigma)}{(k^2-\sigma^2)^3}-\frac{1}{2}\gamma^a \sigma^{cd}R_{cda\mu}
\frac{k^{\mu}}{(k^2-\sigma^2)^2}\right]\;,
\end{eqnarray}
where $\hat{k}=\gamma^{\mu}k_{\mu}$. 

\noindent Note that in the leading order contribution to effective potential
(\ref{2}) $N$ has been factored out.

Making use of Wick rotation and calculating the trace in (\ref{3}), we
obtain
\begin{eqnarray}\label{4}
V'(\sigma)&=&\frac{\sigma}{\lambda}+\frac{2\sigma}{\pi^2}\int_{0}^{\infty}
k^2 dk
\left[-\frac{1}{k^2+\sigma^2}-\frac{R}{12}\frac{1}{(k^2+\sigma^2)^2}
+\frac{2}{9}R\frac{k^2}{(k^2+\sigma^2)^3}\right] \nonumber \\
&=&\frac{\sigma}{\lambda}+\frac{\sigma}{2\pi^{3/2}}\int_{1/{\Lambda^2}}^{\infty}
ds \exp (-s\sigma^2)\left(-s^{-3/2}+\frac{R}{12}s^{-1/2}\right)\;.
\end{eqnarray}
In derivation of eq. (\ref{4}) we have used the expression for
proper-time representation 
\begin{equation}\label{5}
A^{-\nu}=\frac{1}{\Gamma (\nu) }\int_{0}^{\infty}
ds\; s^{\nu-1} e^{-sA}\;,
\end{equation}
and after that, the ultraviolet proper-time cut-off $\Lambda^2$ has
been introduced in the low limit of proper-time integral. Note that in
\cite{b2} another cut-off (introduced as upper limit of momentum
integrals) has been used. 

Performing the integration over $s$ and over $\sigma$, we get:
\begin{eqnarray}\label{6}
V(\sigma)&=&\frac{\sigma^2}{2\lambda}+\frac{1}{6\pi^{3/2}}
\left[\Lambda^3\exp(-\frac{\sigma^2}{\Lambda^2})-
2\sigma^2\Lambda\exp(-\frac{\sigma^2}{\Lambda^2})+2\sqrt{\pi}\sigma^3{\rm
erfc}(\frac{\sigma}{\Lambda})\right.\nonumber \\
&-&\left.\frac{R}{4}\left(\Lambda \exp(-\frac{\sigma^2}{\Lambda^2})-
\sqrt{\pi}\sigma {\rm
erfc}(\frac{\sigma}{\Lambda})\right)\right]\;,
\end{eqnarray}
where ${\rm erfc}(x)= \frac{2}{\sqrt{\pi}} \int^\infty_x e^{-t^2} dt$. Hence,
we obtained the effective potential in $d=3$ NJL model in curved
spacetime using proper-time cut-off.

The renormalized effective potential may be found in the limit $\lambda
\rightarrow \infty$ as following:
\begin{equation}\label{7}
V(\sigma)=\frac{\sigma^2}{2\lambda_{R}}+\frac{|\sigma|^3}{3\pi}+\frac{R|\sigma|}{24\pi}\;,
\end{equation}
where 
\begin{equation}\label{8}
\frac{1}{\lambda_R}=\frac{1}{\lambda}-\frac{\Lambda}{\pi^{3/2}}\;.
\end{equation}

Let us analyse now the phase structure of the potential (\ref{7}). In the
absence of curvature, $R=0$,   for $\lambda_R > 0$ the minimum of the
potential (\ref{7}) is given by $\sigma=0$. Hence, there is no chiral symmetry
breaking. For $\lambda_R < 0$ the chiral symmetry is broken, the
dynamically generated fermion mass is given as

\begin{equation}\label{9}
m \equiv \sigma_{\rm min} = - \frac{\pi}{\lambda_R}\;.
\end{equation}
When the curvature is not zero we find the following picture. Let first
$\lambda_R > 0$. Then the ground state (and dynamically generated mass)
is defined by
\begin{equation}\label{10}
m \equiv \sigma_{\rm min} = - \frac{\pi}{2 \lambda_R} + \frac{1}{2}
\sqrt{\frac{\pi^2}{\lambda^2_R} - \frac{R}{6}}\;.
\end{equation}
One can see that unlike the case of flat space for positive
$\lambda_R$ we have the chiral symmetry breaking. Moreover, as it
follows from \cite{b10} 
we see the possibility of curvature-induced  phase transitions. The
critical curvature is given by $R_c = 0$ (flat space). For negative
curvature, $R< R_c = 0$, we observe the chiral symmetry breaking, while
for positive curvatures symmetry is not broken.

For negative four-fermion coupling constant $\lambda < 0$ we get the
following ground state:
\begin{equation}\label{11}
\sigma_{\rm min} = - \frac{\pi}{2 \lambda_R} - \frac{1}{2}
\sqrt{\frac{\pi^2}{\lambda^2_R} - \frac{R}{6}}\;.
\end{equation}
The critical curvature is defined by the condition: $ R_c = \frac{6
\pi^2}{\lambda_R^2}$. Between $0<R \leq R_c$ the chiral symmetry is
broken.

Now, let us discuss the situation  when NJL model (\ref{1}) is considered
in curved spacetime with external magnetic field. That means that
spinor covariant derivative also contains the electromagnetic
piece. Treating the magnetic field exactly \cite{b9}, one can easily find
the effective potential for the model (\ref{1}). Considering mow NJL model
in curved spacetime with magnetic field, we again work in linear
curvature approximation as above (but making no approximations for an
external magnetic field). Moreover, we take into account only leading
contribution on curvature which doesn't depend on magnetic field,
i.e. the contribution discussed above. (One can show that for not a very
large magnetic field the curvature correction depending explicitly from
magnetic field is not essential). Then, using the eq. (\ref{7}) and the
results of the calculation of  $d=3$ NJL effective potential in an
external magnetic  field \cite{b4,b6}, one can get:
\begin{eqnarray}\label{12}
V&=&\frac{\sigma^2}{2\lambda_{R}}+\frac{|\sigma|^3}{3\pi}+\frac{R|\sigma|}{24\pi}
+\frac{eH}{4\pi^{3/2}}\int_{0}^{\infty}\frac{ds}{s^{3/2}}e^{-s\sigma^2}
\nonumber \\
&\times &\left[\coth(eHs)-\frac{1}{eHs}\right]\;,
\end{eqnarray}
where $H$ is magnetic field and $e$ is electric charge.

The renormalized effective potential (\ref{12}) may be represented as
following
\begin{equation}\label{13}
V=\frac{\sigma^2}{2\lambda_{R}}+\frac{R|\sigma|}{24\pi}+\frac{eH|\sigma|}{2\pi}
-\frac{(2eH)^{3/2}}{2\pi}\zeta\left(-\frac{1}{2},\frac{\sigma^2}{2eH}\right)\;,
\end{equation}
where properties of generalized zeta-function may be found in \cite{b10}.

Working in frames of our approximation and considering also  $eH
\rightarrow 0$, we find that for positive $\lambda_R$
\begin{equation}\label{14}
\sigma_{{\rm min}}\simeq\frac{\lambda_{R}}{\pi}\left(\frac{eH}{2} -\frac{R}{24}\right)  \;.
\end{equation}
Hence, for   $R = 0$ chiral symmetry breaking due to magnetic field
occurs. Negative curvature increases chiral symmetry
breaking. However, positive curvature acts against CSB. On the
critical line of phase diagram
\begin{equation}\label{15}
\frac{R}{12} \simeq eH 
\end{equation}
the restoration of chiral symmetry breaking occurs. Hence, in this
example magnetic field and gravitational field act
in the opposite directions with respect to chiral symmetry breaking. 
Gravity tends to restore the chiral symmetry while magnetic field
tends to break it. On the critical line both effects are
compensated and there is no chiral symmetry breaking

For $\lambda_R < 0,\; eH \rightarrow 0 $ we get
\begin{equation}\label{16}
\sigma_{\rm min}=\sigma_{\rm min}^{H=0}\left[
1+\frac{(eH)^2}{12(\sigma_{\rm min}^{H=0})^4 }
+\ldots \right]\;,
\end{equation}
where $\sigma_{\rm min}^{H=0}$ is given by the expression (\ref{11})
(compare with \cite{b6}).

In summary, we studied $d=3$ NJL model in curved spacetime with
magnetic field. The combined effect of gravitational and magnetic
fields may lead to quite rich phase structure. In particular, magnetic
field always breaks chiral symmetry. However, in curved spacetime with
positive curvature the external gravitational field may lead to
screening of magnetic field and restoration of chiral symmetry as
it is shown above. We expect that this new effect should be relevant for
cosmological applications, for example, in the consideration of GUTs
and SM reformulated as NJL model \cite{b11} in the early Universe with
primordial magnetic fields.
  
SDO would like to thank D. Amati for kind hospitality in SISSA,
Trieste where this work has been completed.

\end{document}